\documentclass[12pt,numbers,sort&compress]{article}

\usepackage{natbib}
\usepackage{amsmath}
\usepackage{amsthm}
\usepackage{amsfonts}
\usepackage{amssymb}
\usepackage[all]{xy}
\usepackage{graphicx}
\usepackage{color}
\usepackage[isu,bf]{caption2}
\newlength{\xtrawidth}
\newlength{\xtraheight}
\def\clap#1{\hbox to 0pt{\hss#1\hss}}

\newenvironment{descriptionlist}{%
\begin{list}%
{}%
{}}%
{\end{list}}
{\everymath{\displaystyle\everymath{}}\array}%
{\endarray}

\newcommand{\eqdef}{%
  \mathrel{\lower.1mm
    \hbox{$\stackrel{\lower.424ex\hbox{\scriptsize def}}{=}$}}
}

\DeclareMathOperator{\Hom}{Hom}
\DeclareMathOperator{\Ext}{Ext}

\DeclareMathOperator{\rank}{rank}

\DeclareMathOperator{\Span}{span}
\DeclareMathOperator{\spec}{spec}

\newcommand{\Z}{\mathbb{Z}}

\newcommand{\IR}{\mathbb{R}}
\newcommand{\C}{\mathbb{C}}
\newcommand{\Q}{\mathbb{Q}}

\newcommand{\CP}[1]{\mathbb{P}^{#1}}
\newcommand{\IP}[1]{\CP{#1}}

\newtheorem{proposition}{Proposition}

\newcommand{\Rep}[1]{\ensuremath{\mathbf{#1}}}
\newcommand{\barRep}[1]{\ensuremath{\overline{\Rep{#1}}}}

\newcommand{\dual}{\ensuremath{\vee}}

\newcommand{\ZZZ}{{\ensuremath{\Z_3\times\Z_3}}}
\newcommand{\B}[1]{\ensuremath{B_{#1}}}
\newcommand{\dP}[1]{\ensuremath{dP_{#1}}}
\newcommand{\Xt}{{\ensuremath{\widetilde{X}}}}

\newcommand{\Osheaf}{\ensuremath{\mathcal{O}}}
\newcommand{\oP}{\ensuremath{\Osheaf_{\CP1}}}
\newcommand{\oB}[1]{\ensuremath{\Osheaf_{\B{#1}}}}
\newcommand{\oXt}{\ensuremath{\Osheaf_{\Xt}}}

\newcommand{\V}[1]{\ensuremath{\mathcal{V}_{#1}}}
\newcommand{\Vt}{{\ensuremath{\widetilde{\V{}}}}}
\newcommand{\W}[1]{{\ensuremath{\mathcal{W}_{#1}}}}

\newcommand{\Fsheaf}{\ensuremath{\mathcal{F}}}
\newcommand{\Lsheaf}{\ensuremath{\mathcal{L}}}
\newcommand{\Msheaf}{\ensuremath{\mathcal{M}}}
\newcommand{\Wsheaf}{\ensuremath{\mathcal{W}}}

\newcommand{\Kcone}{\ensuremath{\mathcal{K}}}

\newcommand{\even}{\ensuremath{\text{ev}}}

\begin{document}
\begin{titlepage}
  \begin{flushright}
    hep-th/0602073
    \\
    UPR 1147-T
  \end{flushright}
  \vspace*{\stretch{1}}
  \begin{center}
     \Huge 
     Stability of the Minimal Heterotic\\ Standard Model Bundle
  \end{center}
  \vspace*{\stretch{2}}
  \begin{center}
    \begin{minipage}{\textwidth}
      \begin{center}
        \large         
        Volker Braun$^{1,2}$, 
        Yang-Hui He$^{3,4}$ and
        Burt A.~Ovrut$^{1}$
      \end{center}
    \end{minipage}
  \end{center}
  \vspace*{1mm}
  \begin{center}
    \begin{minipage}{\textwidth}
      \begin{center}
        ${}^1$ Department of Physics,
        ${}^2$ Department of Mathematics
        \\
        University of Pennsylvania,        
        Philadelphia, PA 19104--6395, USA
      \end{center}
      \begin{center}
        ${}^3$ Merton College, Oxford University,      
        Oxford OX1 4JD, U.K.
      \end{center}
      \begin{center}
        ${}^4$ Mathematical Institute, Oxford, 
        24-29 St.\ Giles', OX1 3LB, U.K.
      \end{center}
    \end{minipage}
  \end{center}
  \vspace*{\stretch{1}}
  \begin{abstract}
    \normalsize 
% -----------------------------------------------------------------------------
    The observable sector of the ``minimal heterotic standard model''
    has precisely the matter spectrum of the MSSM: three families of
    quarks and leptons, each with a right-handed neutrino, and one
    Higgs--Higgs conjugate pair. In this paper, it is explicitly
    proven that the $SU(4)$ holomorphic vector bundle leading to the
    MSSM spectrum in the observable sector is slope-stable.
% -----------------------------------------------------------------------------
  \end{abstract}
  \vspace*{\stretch{5}}
  \begin{minipage}{\textwidth}
    \underline{\hspace{5cm}}
    \centering
    \\
    Email: 
    \texttt{vbraun, ovrut@physics.upenn.edu},
    \texttt{yang-hui.he@merton.ox.ac.uk}.
  \end{minipage}
\end{titlepage}

\tableofcontents

\section{Introduction}
\label{sec:intro}

The $E_8\times E_8$ heterotic string~\cite{Gross:1985fr, Gross:1985rr,
  Horava:1995qa} is, perhaps, the simplest context in which to
construct string compactifications giving rise to a realistic matter
spectrum; that is, three families of quarks/leptons and one (perhaps
several) Higgs--Higgs conjugate pairs without any exotic
representations or any other vector-like pairs. Within the last year,
there has been significant progress in building such
models~\cite{HetSM1, HetSM2, HetSM3, MinimalHetSM, Buchmuller:2005jr,
  Bouchard:2005ag}. In the vacua presented in~\cite{HetSM1, HetSM2,
  HetSM3}, called heterotic standard models, the observable sector has
the MSSM matter spectrum with the addition of one extra pair of Higgs
fields. In~\cite{MinimalHetSM} the number of Higgs pairs was reduced
to one, yielding the exact MSSM matter spectrum in the observable
sector. Hence, the vacua in~\cite{MinimalHetSM} are called ``minimal''
heterotic standard models. The MSSM matter spectrum has been obtained,
in different contexts, in~\cite{Buchmuller:2005jr, Bouchard:2005ag}.

In this paper, we will confine our discussion to the heterotic
standard model vacua presented in~\cite{HetSM1, HetSM2, HetSM3,
  MinimalHetSM}. Their basic construction is as follows. As is well
known, the whole matter content of the standard model, including the
right-handed neutrino~\cite{Fukuda:1998mi, Langacker:2004xy,
  Giedt:2005vx} fits into the $\Rep{16}$ and $\Rep{10}$
representations of $Spin(10)$. To embed this unification of
quarks/leptons into the $E_8\times E_8$ heterotic string, one has to
break the observable sector $E_8$ gauge group appropriately. This can
be done by choosing a suitable gauge instanton~\cite{Donagi:1998xe,
  Donagi:1999gc, Andreas:1999ty, SM-bundle1, DonagiPrincipal,
  Curio:2004pf, Andreas:2003zb, Diaconescu:1998kg} as the vacuum field
configuration on a Calabi-Yau threefold. In particular, an $SU(4)$
instanton leaves a $Spin(10)$ gauge group unbroken~\cite{HetSM1}. The
corresponding rank $4$ vector bundle is constructed via the method of
bundle extensions~\cite{MR1807601, SU5-z2-1, SU5-z2-2}. Of course, the
$Spin(10)$ gauge group must be further broken to a group containing
the standard model gauge group as a factor. The obvious mechanism is
to add Wilson lines~\cite{Witten:1985xc, Sen:1985eb, Breit:1985ud,
  Ibanez:1986tp}, thus breaking $Spin(10)$ directly at the
compactification scale. In particular, we use a $\ZZZ$ Wilson line to
break down to $SU(3)_C\times SU(2)_L\times U(1)_Y\times U(1)_{B-L}$.
In order to do so, the Calabi-Yau threefold must have a large enough
fundamental group~\cite{SM-bundle1, Donagi:2003tb, dP9torusfib,
  Ovrut:2003zj}, that is, it must contain a $\ZZZ$. A Calabi-Yau
threefold whose fundamental group is exactly $\ZZZ$ was constructed
in~\cite{dP9Z3Z3}, and is used in~\cite{HetSM1, HetSM2, HetSM3,
  MinimalHetSM}. The low energy particle spectrum can then be computed
using methods of algebraic geometry as discussed
in~\cite{Green:1987mn, Donagi:2004qk, Donagi:2004ia}.

An important phenomenological aspect of heterotic standard model vacua
is the $U(1)_{B-L}$ factor occurring in the low energy gauge group.
Usual nucleon decay is suppressed in~\cite{HetSM1, HetSM2, HetSM3,
  MinimalHetSM} by a large compactification mass of
$O\big(10^{16}\big)~\text{GeV}$. In addition, these theories exhibit
natural doublet-triplet splitting, thus suppressing proton decay via
dimension five operators. The role of the gauged $U(1)_{B-L}$ symmetry
is to disallow any $\Delta L=1$ and $\Delta B=1$ dimension four terms
that would lead to the disastrous decay of
nucleons~\cite{Nath:2006ut}. Of course, this symmetry must be
spontaneously broken at the order of the electroweak scale. This will
be discussed elsewhere~\cite{future:proton}. Hence, only the usual
Yukawa couplings and a possible Higgs $\mu$-term can occur in the
superpotential at the renormalizable level.  Geometrically, these
couplings are cubic products of cohomology groups and restricted by
classical geometry.  The effect of the elliptic fibration of the
Calabi-Yau threefold on the Yukawa texture was analyzed
in~\cite{Braun:2006me}, and leads to one naturally light quark/lepton
family.

An essential requirement of these vacua is that the holomorphic vector
bundle used in the observable sector be slope-stable. This
guarantees~\cite{MR86h:58038, MR88i:58154} that the associated gauge
connection satisfies the hermitian Yang-Mills equations and, hence,
preserves $\mathcal{N}=1$ supersymmetry. The vector bundles in the
observable sector of~\cite{HetSM1, HetSM2, HetSM3} were shown to be
slope-stable in~\cite{Gomez:2005ii}. In this paper, we present an
analogous proof that the $SU(4)$ vector bundle in the minimal
heterotic standard model~\cite{MinimalHetSM} is, indeed, slope-stable
as well. Thus, the observable sector containing exactly the matter
spectrum of the MSSM is $\mathcal{N}=1$ supersymmetric.

The structure of the hidden sector is less clear. There is a maximal
dimensional subcone (codimension zero) of the K\"ahler cone where the
observable sector bundle is slope-stable and the hidden sector
satisfies the Bogomolov bound. Hence, there is no obstruction to
constructing anomaly free vacua whose hidden sector bundle is
slope-stable. However, we have not explicitly constructed such a
hidden sector bundle. Nor is it entirely clear that this is desirable.
As discussed in~\cite{Kachru:2003aw, Buchbinder:2003pi,
  Buchbinder:2004im}, the necessity to stabilize all moduli at a point
with a small positive cosmological constant~\cite{Riess:1998cb} might
require that the vacuum, in the heterotic case, contain
anti-five-branes. If the moduli can be stabilized for such a
configuration then, for example, a trivial hidden sector bundle (which
is trivially slope-stable) can be chosen.  This issue will be
discussed in detail elsewhere. We note that the slope-stability of
both the observable and hidden sector bundles was proven for the
vacuum in~\cite{Bouchard:2005ag}.

\section{The Calabi-Yau Manifold}
\label{sec:CY}

\subsection{Double Fibration}
\label{sec:fib}

Let us start by describing the underlying Calabi-Yau threefold. We
begin with an elliptic fibration over a rational elliptic ($\dP9$)
surface. Such an elliptic fibration is automatically a fiber product
\begin{equation}
  \Xt \eqdef B_1 \times_{\IP1} B_2
\end{equation}
of two $\dP9$ surfaces $B_1$ and $B_2$. In the following, we always
choose surfaces with suitable $\ZZZ$ automorphisms~\cite{dP9Z3Z3}
yielding a free $\ZZZ$ group action on $\Xt$. There is a commutative
square of projections
\begin{equation}
  \label{eq:CanonicalBundles}
  \vcenter{\xymatrix@!0@C=24mm@R=26mm{
      \dim_\C=3: && & 
      \Big(\Xt,\, K_{\Xt}= \oXt\Big)
      \ar[dr]_{\pi_2}^
      {\displaystyle K_{\Xt|\B2}=\chi_1^2\oXt(\phi)}
      \ar[dl]^{\pi_1}_
      {\displaystyle K_{\Xt|\B1}=\oXt(\phi)}
      \\
      \dim_\C=2: && 
      \Big(\B1,\, K_{\B1}= \oB1(-f) \Big)
      \ar[dr]^{\beta_1}_
      {\displaystyle K_{B_1|\IP1}=\chi_1^2\oB1(f)} & & 
      \Big(\B2,\, K_{\B2}=\chi_1 \oB2(-f) \Big)
      \ar[dl]_{\beta_2}^
      {\displaystyle K_{\B2|\IP1}=\oB2(f)} \\
      \dim_\C=1: && & 
      \Big(\IP1,\, K_{\IP1} = \chi_1 \oP(-2) \Big)
      \,,
  }}
\end{equation}
where $\chi_1$, $\chi_2$ are characters~\cite{HetSM3} of $\ZZZ$
encoding the equivariant action on bundles.

The quotient 
\begin{equation}
  X \eqdef \Xt \Big/ \big(\ZZZ\big)
\end{equation}
is a torus-fibered Calabi-Yau threefold with fundamental group
$\pi_1\big(X\big)=\ZZZ$, which we take to be the base manifold of our
string compactification. However, in practice we work with equivariant
constructions on the universal cover $\Xt$. For a free group action
these descriptions are equivalent.

\subsection{Topology}
\label{sec:topology}

It is important to understand the even cohomology groups
$H^\even\big(X,\Z\big)$, because that is where the Chern classes
live. Rationally, it is clear that
\begin{equation}
  H^\even\Big(\Xt,\Q\Big)^\ZZZ = 
  H^\even\Big(X,\Q\Big)
  \,.
\end{equation}
The degree $2$ invariant integral cohomology of $\Xt$ is
\begin{equation}
  H^2\Big(\Xt,\Z\Big)^\ZZZ = 
  \Span_\Z\big\{ \tau_1,\tau_2,\phi \big\}
  \,.
\end{equation}
We can compare it with the cohomology of $X$ using the quotient map
\begin{equation}
  q:~
  \Xt \to X
  \qquad \Rightarrow \quad
  q^\ast:~
  H^\ast\big(X,\Z\big) \to H^\ast\big(\Xt,\Z\big)
  \,.
\end{equation}
In degree $2$, the image is an index $3$ sub-lattice of
$H^2\big(\Xt,\Z\big)\simeq \Z^3$ generated by $\tau_1-\tau_2$,
$3\tau_1$, $\phi$. In other words, the equivariant line bundles on
$\Xt$ are of the form
\begin{equation}
  \oXt(x_1 \tau_1 + x_2 \tau_2 + x_3 \phi)
  \qquad
  x_1,x_2,x_3 \in \Z
  \,,~
  x_1+x_2\equiv 0 \mod 3    
  \,.
\end{equation}
The products of the degree $2$ generators can easily be determined,
and one finds relations
\begin{equation}
  H^\even\Big( \Xt, \Q \Big)^\ZZZ = 
  \Q\big[ \tau_1,\tau_2,\phi \big]
  \Big/ 
  \left<
    \phi^2
    ,~ 
    \tau_i\phi=3\tau_i^2 
  \right>
  \,.
\end{equation}
Hence, every even degree cohomology class can be written as a
polynomial in $\tau_1$, $\tau_2$, and $\phi$ subject to the relations
$\phi^2=0$ and $\tau_i\phi=3\tau_i^2$.

\section{Visible Bundle}
\label{sec:visible}

\subsection{Construction of the Bundle}
\label{sec:bundledef}

Having presented the Calabi-Yau manifold, we proceed to define a
holomorphic rank $4$ vector bundle on it. First, define equivariant
rank $2$ vector bundles
\begin{subequations}
\begin{align}
  \label{eq:V1def}
  \V1 =&~ 
  \oXt\big( -\tau_1+\tau_2 \big) \otimes
  \pi_1^\ast \big(\W1\big)
  \\
  \label{eq:V2def}
  \V2 =&~ 
  \oXt\big( +\tau_1-\tau_2 \big) \otimes
  \pi_2^\ast \big(\W2\big)
  \,,
\end{align}
\end{subequations}
where $\W1$ and $\W2$ are rank $2$ vector bundles on $\B1$ and $\B2$
which we will define in detail in Section~\ref{sec:serre},
eqns.~\eqref{eq:W1def} and~\eqref{eq:W2def}. Using these, we define
the desired rank $4$ vector bundle $\Vt$ as an extension
\begin{equation}
  \label{eq:Vdef}
  0
  \longrightarrow
  \V1
  \longrightarrow
  \Vt
  \longrightarrow
  \V2
  \longrightarrow
  0
  \,.
\end{equation}
Using the fact that the first Chern class of $\W{i}$ is trivial,
$\wedge^2\W{i}=\oB{i}$, we first remark that
\begin{equation}
  c_1\big( \Vt \big) = 0
  ~\in H^2\Big(\Xt,\Z \Big)^\ZZZ \simeq \Z^3
  \,.
\end{equation}
But we really want an $SU(4)$ bundle on the quotient
$X=\Xt\big/(\ZZZ)$, that is
\begin{equation}
  c_1\Big( \Vt\Big/\big(\ZZZ\big) \Big) = 0
  ~\in H^2\Big(X,\Z \Big) \simeq \Z^3\oplus \Z_3\oplus \Z_3
  \,.  
\end{equation}
The vanishing of the first Chern class including the torsion part
follows from $\wedge^4\Vt=\oXt$, where $\oXt$ stands for the trivial
line bundle with the trivial $\ZZZ$ equivariant group action.

\subsection{Non-Trivial Extensions}
\label{sec:ext}

We defined the rank $4$ bundle $\Vt$ as a generic extension of the
form eq.~\eqref{eq:Vdef}. Clearly, we have to make sure that a
non-trivial extension exists, since the \emph{trivial extension}
$\V1\oplus\V2$ cannot give rise to an irreducible $SU(4)$ instanton.
The space of extensions is
\begin{multline}
  \Ext^1\Big( \V2, \V1 \Big) = 
  H^1\Big( \Xt, \V1\otimes \V2^\dual \Big) 
  = \\ =
  H^1\Big( \Xt, \oXt(-2\tau_1+2\tau_2) \otimes 
  \pi_1^\ast(\W1) \otimes \pi_2^\ast(\W2^\dual) \Big) 
  = \\ =
  H^1\Big( \Xt, 
  \pi_1^\ast\big(\W1\otimes\oB1(-2t)\big) \otimes
  \pi_2^\ast\big(\W2\otimes\oB2(2t)\big) \Big)    
  \,.
\end{multline}
This cohomology group can directly be computed using the Leray
spectral sequence and the push-down eqns.~\eqref{eq:W12tpushdown}
and~\eqref{eq:W22tpushdown}. One obtains 
\begin{equation}
  H^i\Big( \Xt,~ \V1\otimes\V2^\dual \Big) = 
  \begin{cases}
    0
    & i=3    , \\ 
    8 R[\ZZZ]
    & i=2    , \\ 
    4 R[\ZZZ]
    & i=1    , \\ 
    0
    & i=0    .
  \end{cases}
\end{equation}
where $R[\ZZZ]$ stands for the regular representation, that is, the
sum of all $9$ irreducible representations of $\ZZZ$. Of course, only
invariant extensions give rise to equivariant vector bundles $\Vt$.
The invariant subspace is
\begin{equation}
  \Ext^1\Big( \V2, \V1 \Big)^\ZZZ = 
  H^1\Big( \Xt,~ \V1\otimes\V2^\dual \Big)^\ZZZ =
  4
\end{equation}
is indeed non-zero, so suitable extensions do exist.

\subsection{Low-Energy Spectrum}

The low energy particle spectrum is determined through the cohomology
of $\Vt$ and $\wedge^2\Vt$ according to the decomposition
\begin{equation}
  \Rep{248} =
  \big( \Rep{1},\Rep{45} \big) \oplus  
  \big( \Rep{4},\Rep{16} \big) \oplus 
  \big( \barRep{4}, \barRep{16} \big) \oplus 
  \big( \Rep{6},\Rep{10} \big) \oplus
  \big( \Rep{15},\Rep{1} \big)  
\end{equation}
under $E_8\supset SU(4)\times Spin(10)$. It is easy to show that
$H^i\big(\Xt,\Vt\big)=0$ for $i=0,2,3$. Hence a simple index
computation yields
\begin{equation}
  \label{eq:Vcoh}
  H^i\Big( \Xt,~ \Vt \Big) = 
  \begin{cases}
    0
    & i=3    , \\ 
    0
    & i=2    , \\ 
    3 R[\ZZZ]
    & i=1    , \\ 
    0
    & i=0    .
  \end{cases}
\end{equation}
Furthermore, interrelated long exact sequences~\cite{HetSM3} together with
\begin{equation}
  H^\ast\Big( \Xt,~ \wedge^2\V1 \Big) = 
  H^\ast\Big( \Xt,~ \wedge^2\V2 \Big) = 
  0
\end{equation}
yield
\begin{equation}
  H^i\Big( \Xt,~ \wedge^2\Vt \Big) = 
  H^i\Big( \Xt,~ \V1\otimes\V2 \Big) = 
  H^i\Big( \Xt,~ 
  \pi_1^\ast\big(\W1\big) \otimes \pi_1^\ast\big(\W1\big) \Big)   
  \,.
\end{equation}
The latter is easily computed using the push-down formula
eqns.~\eqref{eq:W1pushdown} and~\eqref{eq:W2pushdown} and the Leray
spectral sequence. The result is that
\begin{equation}
  \label{eq:wedge2Vcoh}
  H^i\Big( \Xt,~ \wedge^2\Vt \Big) = 
  H^i\Big( \Xt,~ \V1\otimes\V2 \Big) = 
  \begin{cases}
    0
    & i=3    , \\ 
    \chi_2 \oplus \chi_2^2 \oplus \chi_1\chi_2^2 \oplus \chi_1^2\chi_2
    & i=2    , \\ 
    \chi_2 \oplus \chi_2^2 \oplus \chi_1\chi_2^2 \oplus \chi_1^2\chi_2
    & i=1    , \\ 
    0
    & i=0    .
  \end{cases}
\end{equation}
Finally, the $\ZZZ$ group action on the cohomology is tensored with
the Wilson line, and every state that is not invariant under the
combined action is projected out. The regular representations in
eq.~\eqref{eq:Vcoh} yield $3$ full generations of quarks and leptons,
each with a right-handed neutrino. More interesting is the Wilson line
action on the $\Rep{10}$ of $Spin(10)$, which potentially could lead
to exotic color triplets (``triplet Higgs''). We chose the Wilson line
such that
\begin{equation}
  \Rep{10}= 
  \Big[
  \chi_2^2\big(\Rep{1},\Rep{2},3,0\big) \oplus 
  \chi_1^2\chi_2^2\big(\Rep{3},\Rep{1},-2,-2\big)
  \Big]
  \oplus 
  \Big[
  \chi_2\big(\Rep{1},\barRep{2},-3,0\big) \oplus 
  \chi_1\chi_2\big(\barRep{3},\Rep{1},2,2\big)
  \Big]
  \label{eq:10Wilson}
\end{equation}
under the decomposition 
\begin{equation}
  \label{eq:Spin10break}
  Spin(10) \supset 
  SU(3)_C\times SU(2)_L \times U(1)_Y \times U(1)_{B-L} \times \ZZZ
  \,.
\end{equation}
Combining eqns.~\eqref{eq:10Wilson} and~\eqref{eq:wedge2Vcoh}, we see
that one vector-like pair of Higgs survives the $\ZZZ$ quotient while
all color triplets are projected out.

\section{Slope-Stability}
\label{sec:stability}

\subsection{Conditions for Stability}

We now proceed and show that the K\"ahler class $\omega \in
H^2\big(\Xt,\IR\big)$ can be chosen such that the visible sector
vector bundle $\Vt$, eq.~\eqref{eq:Vdef}, is
equivariantly\footnote{$\Vt$ being equivariantly stable is the same as
  $\Vt/\big(\ZZZ\big)$ being stable. For the remainder of this
  section, everything is equivariant.} slope-stable. That means that
for all reflexive sub-sheaves $\Fsheaf\hookrightarrow \Vt$, the slope
\begin{equation}
  \label{eq:slope}
  \mu(\Fsheaf) \eqdef \frac{1}{\rank \Fsheaf} 
  \int_\Xt c_1(\Fsheaf) \wedge \omega^2
\end{equation}
is negative,
\begin{equation}
  \mu\big(\Fsheaf\big) < \mu\big(\Vt\big) = 0
\end{equation}
The easiest way to prove this is to derive a set of sufficient
inequalities for the K\"ahler class $\omega$, and then to find a
common solution~\cite{Gomez:2005ii}. We note that they are not always
necessary, that is, the inequalities are not sharp.

For example, consider only $\V1$ defined by eqns.~\eqref{eq:V1def},
\eqref{eq:W1def}. Let $\Lsheaf$ be any sub-line bundle,
that is
\begin{equation}
  \label{eq:V1sesL}
  \vcenter{\xymatrix{
      0 \ar[r] &
      \chi_1 \oXt(-\tau_1+\tau_2-\phi) \ar[r]^-{u} &
      \V1
      \ar[r]^-{v} &
      \chi_1^2 \oXt(-\tau_1+\tau_2+\phi) \otimes \pi_1^\ast I_3
      \ar[r] &
      0
      \\
      & & \Lsheaf \ar[u]_{i} \ar[ur]_{v \circ i} \ar@{-->}[ul]^{w}
  }}
\end{equation}
The composition $v\circ i$ either vanishes or not. We distinguish the
two cases:
\begin{descriptionlist}
\item[$v\circ i=0$:] There exists a non-zero map 
  \begin{equation}
    w:~\Lsheaf\to\chi_1 \oXt(-\tau_1+\tau_2-\phi)
  \end{equation}
  such that $i=u\circ w$.
\item[$v\circ i\not=0$:] There exists a non-zero map 
  \begin{equation}
    v\circ i:~\Lsheaf\to\chi_1^2 \oXt(-\tau_1+\tau_2+\phi)
  \end{equation}
  whose image vanishes at the codimension two locus where $\pi_1^\ast
  I_3$ vanishes.
\end{descriptionlist}
The existence of these maps restricts the line bundle $\Lsheaf$. Now
if $\Vt$ is stable, then all these line bundles $\Lsheaf$ must be of
negative slope, $\mu(\Lsheaf)<0$. We only have to check this
inequality for the $\Lsheaf$ of largest slope, and these form a finite
set (see Appendix~\ref{sec:sublinebundle}):
\begin{descriptionlist}
\item[$v\circ i=0$:] 
  \begin{equation}
    \Lsheaf=\oXt(-\tau_1+\tau_2-\phi)
    \,.
  \end{equation}
\item[$v\circ i\not=0$:] The composition $v\circ i$ cannot be an
  isomorphism, since that would split the short exact sequence
  eq.~\eqref{eq:V1sesL}. Hence, $\Lsheaf$ can only be a proper
  sub-line bundle, and those of largest slope are 
  \begin{multline}
    \Big\{
    \oXt(-\tau_1+\tau_2)
    ,\,
    \oXt(-4\tau_1+\tau_2+2\phi)
    ,\,
    \oXt(-3\tau_1+\phi)
    ,\,
    \\
    \oXt(-2\tau_1-\tau_2+\phi)
    ,\,
    \oXt(-\tau_1-2\tau_2+2\phi)
    \Big\}
    \,.
  \end{multline}
  The first line bundle $\oXt(-\tau_1+\tau_2)$ actually has the same
  fiber degrees (coefficients of $\tau_1$ and $\tau_2$) as the range
  of $v\circ i$. Because of the push-down formula eq.~\eqref{eq:I3pushdown}, 
  the largest such sub-line bundle whose image vanishes at $\pi_1^\ast
  I_3$ is actually
  \begin{equation}
    \oXt(-\tau_1+\tau_2+\phi) \,\otimes\,
    \pi_1^\ast\circ\beta_1^\ast\Big( \oP(-3) \Big)
    = 
    \oXt(-\tau_1+\tau_2-2\phi)
    \,.
  \end{equation}
  Therefore, the possible line bundles $\Lsheaf$ of largest slope are
  \begin{multline}
    \Lsheaf \in 
    \Big\{
    \oXt(-\tau_1+\tau_2-2\phi)
    ,\,
    \oXt(-4\tau_1+\tau_2+2\phi)
    ,\,
    \oXt(-3\tau_1+\phi)
    ,\,
    \\
    \oXt(-2\tau_1-\tau_2+\phi)
    ,\,
    \oXt(-\tau_1-2\tau_2+2\phi)
    \Big\}
    \,.
  \end{multline}
\end{descriptionlist}
Similarly, one obtains a finite set of potentially destabilizing
sub-line bundles of $\V2$. 

Now to prove~\cite{Gomez:2005ii} stability of $\Vt$, it suffices to
show that
\begin{itemize}
\item Sub-line bundles of $\Vt$ have negative slope.
\item Rank $2$ sub-bundles have negative slope. A sufficient criterion
  is that $\wedge^2\V1$ has negative slope and that proper sub-line
  bundles of $\wedge^2\V2$ are of negative slope.
\item Rank $3$ sub-bundles (reflexive sheaves) have negative slope
  $\Leftrightarrow$ sub-line bundles of $\Vt^\dual$ have negative
  slope.
\end{itemize}
This gives a finite set of line bundles which have to have negative
slope. One obtains
\begin{proposition}
  \label{prop:stable}
  If all line bundles $\oXt(a_1\tau_1+a_2\tau_2+b\phi)$ with
  \begin{multline}
    (a_1,a_2,b) \in \Big\{
    (-1, -2,  2),\, 
    ( 2, -2, -1),\, 
    ( 2, -5,  1),\, 
    (-4,  1,  2),\, 
    (-1,  1, -1),\, 
    \\
    (-2,  2,  0),\, 
    (-2, -1,  2),\,
    ( 1, -4,  2),\, 
    ( 1, -1, -1)
    \Big\}
  \end{multline}
  have negative slope, then the vector bundle $\Vt$,
  eq.~\eqref{eq:Vdef}, is equivariantly stable.
\end{proposition}

\subsection{K\"ahler Cone Substructure}
\label{sec:kahlersub}

The K\"ahler cone, that is the set of possible K\"ahler classes,
is~\cite{Gomez:2005ii}
\begin{equation}
  \label{eq:Kcone}
  \Kcone 
  \eqdef
  \Big\{ x_1\tau_1 + x_2\tau_2 + y\phi
  \,\Big|\, x_1, x_2, y >0
  \Big\}
  ~\subset
  H^2\big(\Xt,\IR\big)=\left<\tau_1,\tau_2,\phi\right>_\IR
  \,.
\end{equation}
The slope eq.~\eqref{eq:slope} of a line bundle obviously depends
quadratically on the K\"ahler parameters $x_1$, $x_2$, $y$, and can be
computed~\cite{Gomez:2005ii} to be
\begin{equation}
  \mu\Big( \oXt(a_1 \tau_1 + a_2 \tau_2 + b \phi) \Big) 
  = 
  3 (x_1 x_2 + 6 y) (a_1 x_2 + a_2 x_1) + 
  x_1 x_2 (3 a_1+ 3 a_2 + 18 b)
  \,.
\end{equation}
Therefore, according to Proposition~\ref{prop:stable} the vector
bundle $\Vt$ is stable if the inequalities
\begin{equation}
  \label{eq:inequalities}
  \begin{array}{l@{\,=\,}c@{\,<\,0}c}
    \mu\big(\oXt(-\tau_1 -2\tau_2 +2\phi)\big) &
    18 x_1 x_2-6 x_1^2-3 x_2^2-18y x_2-36y x_1 
    & \\
    \mu\big(\oXt( 2\tau_1 -2\tau_2 -\phi)\big) &
    -6 x_1^2+6 x_2^2+36y x_2-36y x_1-18 x_1 x_2
    & \\
    \mu\big(\oXt( 2\tau_1 -5\tau_2 +\phi)\big) &
    -15 x_1^2+6 x_2^2+36y x_2-90y x_1
    & \\
    \mu\big(\oXt(-4\tau_1 +\tau_2 +2\phi)\big) &
    18 x_1 x_2+3 x_1^2-12 x_2^2-72y x_2+18y x_1
    & \\
    \mu\big(\oXt(-\tau_1 +\tau_2 -\phi)\big) &
    3 x_1^2-3 x_2^2-18y x_2+18y x_1-18 x_1 x_2
    & \\
    \mu\big(\oXt(-2\tau_1 +2\tau_2 )\big) &
    6 x_1^2-6 x_2^2-36y x_2+36y x_1
    & \\
    \mu\big(\oXt(-2\tau_1 -\tau_2 +2\phi)\big) &
    18 x_1 x_2-3 x_1^2-6 x_2^2-36y x_2-18y x_1
    & \\
    \mu\big(\oXt( \tau_1 -4\tau_2 +2\phi)\big) &
    18 x_1 x_2-12 x_1^2+3 x_2^2+18y x_2-72y x_1
    & \\
    \mu\big(\oXt( \tau_1 -\tau_2 -\phi)\big) &
    -3 x_1^2+3 x_2^2+18y x_2-18y x_1-18 x_1 x_2
    & \\
  \end{array}
\end{equation}
are simultaneously satisfied. 

It is easy to see that there are many solutions. For example, the
K\"ahler class
\begin{equation}
  \label{eq:omega}
  \omega = 3\Big( 2 \tau_1 + 3 \tau_2 + \phi \Big)
  ~\in H^2\Big(\Xt,\IR\Big)
\end{equation}
satisfies all the inequalities eq.~\eqref{eq:inequalities}, the slopes
being $-621$, $-378$, $-702$, $-1512$, $-1269$, $-594$, $-918$, $-27$,
and $-675$, respectively. The overall factor of $3$ in
eq.~\eqref{eq:omega} is not essential, but included to make it a
$\ZZZ$-equivariant integral cohomology class. In other words, the
class is actually primitive in the integral cohomology of the quotient
$X=\Xt/(\ZZZ)$. Of course, in string theory the K\"ahler form is not
quantized. As usual, the radial part of the K\"ahler class, that is,
the overall volume, does not matter for the stability of vector
bundles. We conclude from eq.~\eqref{eq:omega} that the set
\begin{equation}
  \Kcone^s\subset \Kcone \subset H^2\big(\Xt,\IR\big)
\end{equation}
of K\"ahler classes that make all slopes of the line bundles in
Proposition~\ref{prop:stable} negative is not empty. Therefore, the
solution set $\Kcone^s$ of the strict inequalities
eq.~\eqref{eq:inequalities} must be a maximal-dimensional subcone of
the K\"ahler cone $\Kcone$.
\begin{figure}[htb]
  \centering
  \input{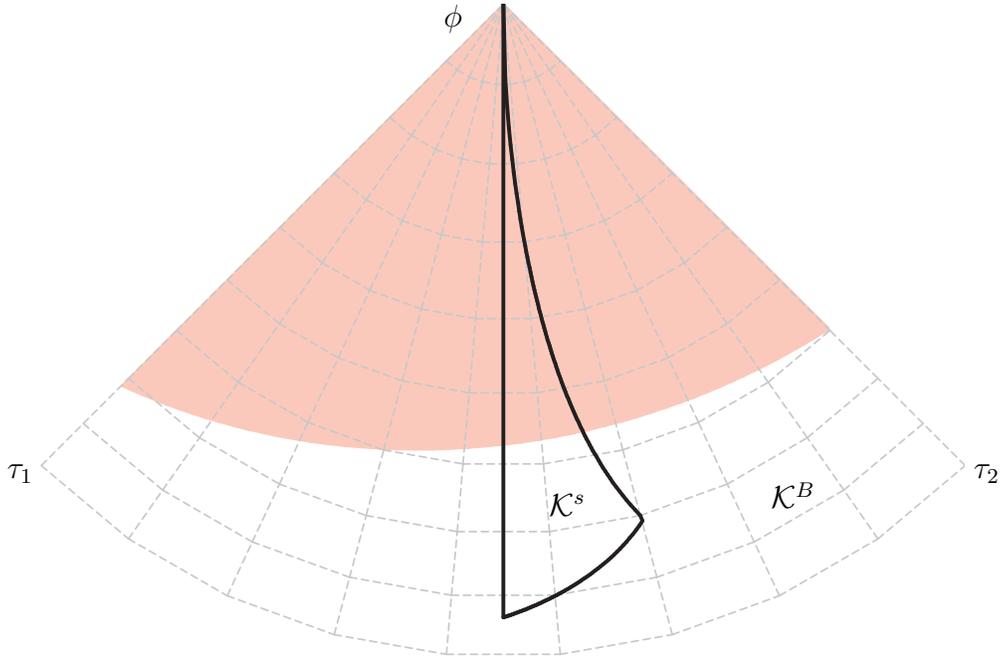}
  \caption{Map projection of the unit sphere intersecting the K\"ahler
    cone, that is, the positive octant in $H^2\big(\Xt,\IR\big)\simeq
    \IR^3$. The rank $4$ bundle $\Vt$ is stable inside the black
    triangular region $\Kcone^s$. In the white region $\Kcone^B$ the
    Bogomolov inequality allows an $\mathcal{N}=1$ hidden sector, see
    Section~\ref{sec:hidden}.}
  \label{fig:stable}
\end{figure}
Note that all cones have their tip at the origin $0\in
H^2\big(\Xt,\IR\big)\simeq \IR^3$. Hence, we can draw a
$2$-dimensional ``star map'' of these cones as they are seen by an
observer at the origin. This is depicted in Figure~\ref{fig:stable}.
One observes that the boundary of the set $\Kcone^s$ is roughly
triangular. On the right hand side in Figure~\ref{fig:stable}, it is
bounded by two curved but smooth faces. Those bounds are an artifact
of our proof, and are merely sufficient but not necessary conditions.
Although it is in general difficult to determine the precise subcone
of the K\"ahler cone where $\Vt$ is stable, one expects it to extend
even further to the right. On the other hand, the flat face of
$\Kcone^s$ at the left in Figure~\ref{fig:stable} is a boundary
saturating a necessary \emph{and} sufficient inequality. It is
precisely the locus where the slope of $\V1$ changes sign, and if one
crosses this line then $\mu\big(\V1\big)>0$ becomes a destabilizing
sub-bundle of $\Vt$, see eq.~\eqref{eq:Vdef}. The interpretation is
analogous to the picture of D-branes as complexes; this boundary of
$\Kcone^s$ is a line of marginal stability. To its right, the bound
state $\Vt$ of $\V1$ and $\V2$ is stable. To its left, the reversed
bound state
\begin{equation}
  \label{eq:Vrevdef}
  0
  \longrightarrow
  \V2
  \longrightarrow
  \Vt_\text{rev}
  \longrightarrow
  \V1
  \longrightarrow
  0
  \,.
\end{equation}
is stable. Using the same methods as above, it is easy to see that
$\Vt_\text{rev}$ is indeed stable in a subcone of $\Kcone$ extending
to the left of $\Kcone^s$. Although reversing the short exact sequence
potentially alters the cohomology groups, it turns out that $\Vt$ and
$\Vt_\text{rev}$ give rise to the same low energy spectrum. 

To summarize, the observable sector vector bundle $\Vt$ is
slope-stable with respect to any K\"ahler class $\omega$ in a
$3$-dimensional subcone $\Kcone^s$ of the $3$-dimensional K\"ahler
cone $\Kcone$. The region $\Kcone^s$ is show explicitly in
Figure~\ref{fig:stable}. By working harder to strengthen
Proposition~\ref{prop:stable} or by making small changes to the vector
bundle it will be possible to enlarge that fraction of the K\"ahler
cone.

\section{Hidden Sector}
\label{sec:hidden}

Although not the main topic of this paper, in this section we will
briefly discuss the hidden sector. Denote by $\Vt'$ the holomorphic
vector bundle of the hidden sector. For simplicity, we will assume
that $c_1\big(\Vt'\big)=0$, that is, the hidden sector contains an
$SU(n)$ gauge instanton. Given the tangent bundle $T\Xt$ of the
Calabi-Yau threefold and the observable sector bundle $\Vt$, anomaly
cancellation imposes the constraint
\begin{equation}
  c_2\big(\Vt'\big) 
  =
  c_2\big(T\Xt\big) -
  c_2\big(\Vt\big) -
  [C_5] 
  \,.
\end{equation}
Here, $[C_5]$ is the curve class on which five-branes are wrapped. For
simplicity, let us assume that $[C_5]=0$ (both weakly and strongly
coupled heterotic string). Then, using
\begin{equation}
  c_2\big(T\Xt\big)
  = 
  12\big( \tau_1^2+\tau_2^2 \big)
  \,,\qquad
  c_2\big(\Vt\big)   
  = 
  \tau_1^2 + 4 \tau_2^2 + 4\tau_1\tau_2
  \,,
\end{equation}
it follows that
\begin{equation}
    c_2\big(\Vt'\big)
    =
    11 \tau_1^2 + 8 \tau_2^2 - 4 \tau_1 \tau_2
    =
    \Big( 3\tau_1^2\Big) + 4 \Big( \tau_1^2+\tau_2^2 \Big) 
    - 4 \Big( \tau_1\tau_2-\tau_1^2-\tau_2^2 \Big)
    \,.    
\end{equation}
Note that $c_2\big(\Vt'\big)$ is neither effective nor antieffective,
the terms in brackets being pull-backs of effective curves on $X$. If
$\Vt'$ is a slope-stable vector bundle with respect to a K\"ahler
class $\omega$, then it must satisfy the Bogomolov
inequality~\cite{MR522939}
\begin{equation}
  \label{eq:bogomolov}
  \int_\Xt c_2\big(\Vt'\big) \wedge \omega > 0
  \,.
\end{equation}
Using the parametrization of $\omega$ in eq.~\eqref{eq:Kcone}, we see
that
\begin{equation}
  \begin{split}
    \int_\Xt  c_2\big(\Vt'\big) \wedge \omega
    =&~ 
    \int_\Xt 
    \Big( 11 \tau_1^2 + 8 \tau_2^2 - 4 \tau_1 \tau_2 \Big)
    \wedge
    \Big( x_1\tau_1+x_2\tau_2+y\phi \Big) 
    = \\ =&~ 
    \int_\Xt     
    \Big( 4 x_1 + 7 x_2 - 12 y \Big) \tau_1^2\tau_2
    =
    3 \Big( 4 x_1 + 7 x_2 - 12 y \Big)
    \,.
  \end{split}
\end{equation}
Therefore, the Bogomolov inequality is satisfied for any K\"ahler
class for which
\begin{equation}
  \label{eq:bogoineq}
  4 x_1 + 7 x_2 - 12 y > 0
  \,.
\end{equation}
This defines a $3$-dimensional cone in the K\"ahler moduli space which
we denote by $\Kcone^B$.  The subcone $\Kcone^B$ is shown as the white
region in Figure~\ref{fig:stable}. Its complement, where
eq.~\eqref{eq:bogoineq} is violated, is drawn in pink.  Note that the
K\"ahler class eq.~\eqref{eq:omega} for which the observable sector
vector bundle was proven to be stable also satisfies
eq.~\eqref{eq:bogoineq}. Hence,
\begin{equation}
  \Kcone^s \cap \Kcone^B \not= \emptyset
  \,.
\end{equation}
Since both $\Kcone^s$ and $\Kcone^B$ are open (solutions of strict
inequalities), their non-empty intersection is automatically a
maximal-dimensional subcone of the K\"ahler cone.  It follows that
both $\Vt$ and $\Vt'$ can, in principle, be slope-stable with respect
to a K\"ahler class in $\Kcone^s\cap\Kcone^B$. Often, the Bogomolov
inequality is the only obstruction to finding stable bundles. However,
we have not explicitly constructed such a hidden sector bundle.

\section{Serre Construction}
\label{sec:serre}

\subsection{General Construction}
\label{sec:general}

In this section, we are going to construct two $SU(2)$ vector bundles
$\W1$ and $\W2$ on the $\dP9$ surfaces $\B1$ and $\B2$,
respectively. They are defined as extensions of the form
\begin{subequations}
\begin{gather}
  \label{eq:W1def}
  0 
  \longrightarrow
  \chi_1 \oB1(-f) 
  \longrightarrow
  \W1
  \longrightarrow
  \chi_1^2 \oB1(f) \otimes I_3
  \longrightarrow
  0
  \\
  \label{eq:W2def}
  0 
  \longrightarrow
  \chi_2^2 \oB2(-f) 
  \longrightarrow
  \W2
  \longrightarrow
  \chi_2 \oB2(f) \otimes I_6
  \longrightarrow
  0
\end{gather}
\end{subequations}
with the ideal sheaves $I_3$ and $I_6$ defined in
Subsection~\ref{sec:ideal}. If they satisfy the Cayley-Bacharach
property, then $\W1$ and $\W2$ are rank $2$ vector bundles for generic
extensions. We check this in Subsection~\ref{sec:CB}.

Note that the determinant line bundles are trivial by construction,
that is
\begin{equation}
  \wedge^2 \W1 = \oB1
  \,,\qquad
  \wedge^2 \W2 = \oB2
  \,.
\end{equation}
Therefore, the bundles are self-dual,
\begin{equation}
  \big(\W1\big)^\dual = \W1
  \,,\qquad
  \big(\W2\big)^\dual = \W2
  \,.
\end{equation}

\subsection{Ideal Sheaves}
\label{sec:ideal}

Let $p_1$, $p_2$, $p_3$ be the singular points of the $3I_1$ Kodaira
fibers in $\B1\to\IP1$. Similarly, let $q_1$, $q_2$, $q_3$ be the
singular points of the $3I_1$ Kodaira fibers in $\B2\to\IP1$.  Recall
that $\ZZZ$ is generated by $g_1$ and $g_2$, where $g_1$ acts on the
base $\IP1$ and $g_2$ does not (it is a translation along the elliptic
fiber). The $\ZZZ$ characters are defined via
\begin{equation}
\begin{aligned}
  \chi_1(g_1) &= \omega 
  & \qquad
  \chi_1(g_2) &= 1
  \\
  \chi_2(g_1) &= 1
  &
  \chi_2(g_2) &= \omega 
  \,,
\end{aligned}
\end{equation}
Note that the points $p_i$ and $q_j$ are $g_2$-fixed points, and that
$g_2$ acts as $\chi_2\oplus\chi_2^2$ on the tangent spaces
$T_{p_i}\B1$ and $T_{q_j}\B2$. First, we define the ideal sheaf $I_3$
as
\begin{equation}
  \label{eq:I3def}
  0 
  \longrightarrow
  I_3
  \longrightarrow
  \oB1
  \longrightarrow
  \bigoplus_{i=1,2,3} \Osheaf_{p_i}
  \longrightarrow
  0  
  \,.
\end{equation}
Furthermore, define for any $G_2\simeq \Z_3$ fixed point $p$ the
subscheme $Z(p)$ as the point $p$ and its first derivative in
$\chi_2^2$-direction. In local coordinates $(x,y)\in \C^2$, this
$\Z_3$ group acts as
\begin{equation}
  g_2(x,y) = \Big( \chi_2(g_2)\, x, \chi_2^2(g_2)\, y \Big)
  = 
  \big( \omega x, \omega^2 y \big)
  \,, \quad
  \omega \eqdef e^{\frac{2\pi i}{3}}
\end{equation}
and the scheme $Z(p)$ is 
\begin{equation}
  Z(p) = \spec\Big( \C[x,y]\big/ \left<x, y^2\right> \Big)
  \,.
\end{equation}
Define the ideal sheaf $I_6$ as the sheaf of functions vanishing at
$\Z(q_1)$, $Z(q_2)$, and $Z(q_3)$. That is, 
\begin{equation}
  \label{eq:I6def}
  0 
  \longrightarrow
  I_6
  \longrightarrow
  \oB2
  \longrightarrow
  \bigoplus_{i=1,2,3} \Osheaf_{Z(q_i)}
  \longrightarrow
  0  
  \,.
\end{equation}
In other words, $I_6$ are the functions vanishing at $q_i$ and whose
first derivative in the $\chi_2^2$ direction vanishes.

\subsection{Cayley-Bacharach Property}
\label{sec:CB}

Recall the Cayley-Bacharach property for an extension
\begin{equation}
  \label{eq:CB}
  0 
  \longrightarrow
  \Lsheaf 
  \longrightarrow
  \Wsheaf
  \longrightarrow
  \Msheaf \otimes I_n
  \longrightarrow
  0
\end{equation}
of line bundles $\Lsheaf$, $\Msheaf$ and ideal sheaf $I_n$ of $n$
points on a surface $B$. It has the Cayley-Bacharach property if the
sections
\begin{equation}
  \label{eq:CBsec}
  s 
  ~\in 
  H^0\Big(B,~ \Lsheaf^\dual \otimes \Msheaf \otimes K_B \Big) 
\end{equation}
vanishing at $n-1$ points of the ideal sheaf automatically vanish at
the $n$-th point. The Cayley-Bacharach property implies that $\Wsheaf$
is generically a rank $2$ vector bundle.

First, let us check that $\W1$, eq.~\eqref{eq:W1def}, has the
Cayley-Bacharach property. The sections in question are
\begin{equation}
  \begin{split}
    s_1 
    ~\in&~
    H^0 \Big( B_1,~
    \oB1(-f)^\dual \otimes \oB1(f) \otimes K_{\B1} \Big) =
    \\ &~=
    H^0 \Big( B_1,~ \oB1(f) \Big)
    = 
    H^0 \Big( \IP1,~ \oP(1) \Big)
    \,.
\end{split}
\end{equation}
Furthermore, the ideal sheaf $I_3$ vanishes at $3$ points in $3$
different fibers. But a section of $\oB1(f)$ can only vanish at one
fiber, or it is identically zero. Hence, a section $s_1$ vanishing at
$2$ of the $3$ points vanishes automatically at the $3$-rd, and the
Cayley-Bacharach property holds.  The extension $\W2$,
eq.~\eqref{eq:W2def}, satisfies Cayley-Bacharach analogously.
Therefore, $\W1$ and $\W2$ are rank $2$ vector bundles.

\subsection{Push-Down Formulae}
\label{sec:Wpushdown}

To compute the cohomology groups of vector bundles, we always utilize
the Leray spectral sequence. For that, we need to know the push-down
of all bundles involved. 

First, consider the ideal sheaves. A standard application of the long
exact sequence for the push-down to eq.~\eqref{eq:I3def} immediately
yields
\begin{equation}
  \label{eq:I3pushdown}
  \beta_{1\ast} \big(I_3\big) = \oP(-3)
  \,, \qquad
  R^1 \beta_{1\ast} \big(I_3\big) = \chi_1 \oP(-1)  
  \,.
\end{equation}
For the push-down of $I_6$ defined in eq.~\eqref{eq:I6def}, first note
that according to the definition of $Z(q_i)$ the push-down of the
skyscraper sheaves are
\begin{equation}
  \beta_{2\ast} \Osheaf_{Z(q_i)} = 
  \Osheaf_{\beta_{2\ast}(q_i)} \oplus 
  \chi_2^2 \Osheaf_{\beta_{2\ast}(q_i)}
  \,.
\end{equation}
The long exact sequence for the push-down contains a non-zero
coboundary map which can be computed as in~\cite{HetSM3}. One finds
that
\begin{equation}
  \label{eq:I6pushdown}
  \beta_{2\ast} \big(I_6\big) = \oP(-3)
  \,, \qquad
  R^1 \beta_{2\ast} \big(I_6\big) = \oP(-1) \oplus 
  \left[ \bigoplus_{i=1}^3 \chi_2^2 \Osheaf_{\beta_2(q_i)} \right]
  \,.
\end{equation}

Using the push-down of the ideal sheaves, we find the long exact
sequence
\begin{equation}
  \label{eq:W1les}
  \vcenter{\xymatrix@R=10pt@M=4pt@H+=22pt{
      0 \ar[r] & 
      \chi_1 \oP(-1)
      \ar[r] &
      \beta_{1\ast} \big( \W1 \big)
      \ar[r] &
      \chi_1^2 \oP(-2) 
      \ar`[rd]^<>(0.5){\delta}`[l]`[dlll]`[d][dll] & 
%      \ar`[rd]`[l]`[dlll]`[d][dll] & 
      \\
      & 
      \chi_1^2 \oP(-2)
      \ar[r] &
      R^1\beta_{1\ast} \big( \W1 \big)
      \ar[r] &
      \oP
      \ar[r] &
      0
      \,.
    }}
\end{equation}
From the discussion is Subsection~\ref{sec:CB} we know that
$\W1=\W1^\dual$ is a vector bundle, that is, it satisfies the relative
duality for vector bundles
\begin{equation}
  R^1\beta_{1\ast} \big(\W1\big) = \Big( \beta_{1\ast} \big(\W1\big) 
  \otimes K_{\B1|\IP1} \Big)^\dual   
  \,.
\end{equation}
This uniquely fixes the coboundary map $\delta$ to be an isomorphism,
and one obtains
\begin{equation}
  \label{eq:W1pushdown}
  \begin{split}
    \beta_{1\ast} \W1
    =&~
    \chi_1 \oP(-1)
    \,,
    \\
    R^1 \beta_{1\ast} \W1
    =&~
    \oP
    \,.
  \end{split}
\end{equation}
The coboundary map in the analogous push-down of $\W2$ is zero for
trivial reasons. We find that
\begin{equation}
  \label{eq:W2pushdown}
  \begin{split}
    \beta_{2\ast} \W2
    =&~
    \chi_2^2 \oP(-1) \oplus \chi_2 \oP(-2)
    \,,
    \\
    R^1 \beta_{2\ast} \W2
    =&~
    \chi_2^2 \oP(1) \oplus \chi_2 \oP
    \,.
  \end{split}
\end{equation}

Finally, we need the push-down of $\W{i}\otimes\oB{i}(2t)$. These are
simpler to compute since the fiber degrees are large, so
$R^1\beta_{i\ast}$ vanishes. First, the push-down of the ideal sheaves
twisted by $\oB{i}(2t)$ is
\begin{subequations}
\begin{align}
  \beta_{1\ast} \Big( I_3\otimes\oB1(2t) \Big) =&~ 
  3 \oP \oplus 3 \oP(-1)
  \,, &
  R^1\beta_{1\ast} \Big( I_3\otimes\oB1(2t) \Big) =&~ 0
  \,, \\
  \beta_{2\ast} \Big( I_6\otimes\oB2(2t) \Big) =&~ 
  6 \oP(-1)
  \,, &
  R^1\beta_{2\ast} \Big( I_6\otimes\oB2(2t) \Big) =&~ 0
  \,.
\end{align}
\end{subequations}
The push-down long exact sequence for $\W1$, $\W2$
splits~\cite{HetSM3}, and we obtain
\begin{equation}
\begin{split}
  \label{eq:W12tpushdown}
  \beta_{1\ast} \Big( \W1\otimes\oB1(2t) \Big) =&~
  6 \oP(-1) \oplus 3 \oP \oplus 3 \oP(1)
  \,, \\
  R^1\beta_{1\ast} \Big( \W1\otimes\oB1(2t) \Big) =&~ 0
  \,, \\
\end{split}
\end{equation}
and
\begin{equation}
\begin{split}
  \label{eq:W22tpushdown}
  \beta_{2\ast} \Big( \W2\otimes\oB2(2t) \Big) =&~
  6 \oP(-1) \oplus 6\oP
  \,, \\
  R^1\beta_{2\ast} \Big( \W2\otimes\oB2(2t) \Big) =&~ 0
  \,.
\end{split}
\end{equation}
The push-down for $\W{i}\otimes\oB{i}(-2t)$ can be obtained by
relative duality.

\section*{Acknowledgements}

We are grateful to E.~Buchbinder, R.~Donagi, P.~Langacker, B.~Nelson
and D.~Waldram for enlightening discussions. This research was
supported in part by cooperative research agreement DE-FG02-95ER40893
with the U.~S.~Department of Energy and an NSF Focused Research Grant
DMS0139799 for ``The Geometry of Superstrings''.  Yang Hui-He is
supported in part by the FitzJames Fellowship at Merton College,
Oxford.

\appendix

\section{Line Bundles}
\label{sec:sublinebundle}

By elementary computation of $\Hom(\Lsheaf,\oXt)$, one can easily see
that every equivariant sub-line bundle $\Lsheaf$ of $\oXt$ is
\begin{equation}
  \label{eq:oXtsublb}
  \oXt(-\phi)
  ,\,
  \oXt(-3 \tau_1 + \phi)  
  ,\,
  \oXt(-2 \tau_1 -\tau_2)  
  ,\,
  \oXt(-\tau_1-2 \tau_2)  
  ,\,
  \oXt(-3 \tau_2 + \phi)  
\end{equation}
or a sub-line bundle thereof. Since a sub-line bundle of a line bundle
always has smaller slope, the equivariant sub-line bundles of $\oXt$
of largest slope are those listed in eq.~\eqref{eq:oXtsublb}.

\bibliographystyle{JHEP} \renewcommand{\refname}{Bibliography}
\addcontentsline{toc}{section}{Bibliography} \bibliography{main}

\end{document}